\documentstyle[aps,preprint]{revtex}
\begin{document}
\draft
\title
{Search for the nondimerized quantum nematic phase in the spin-1 chain}
\author{G.\ F\'ath and J.\ S\'olyom\cite{byline}}
\address{Research Institute for Solid State Physics \\
               H-1525 Budapest, P.\ O.\ Box 49, Hungary}
\date{1.\ October 1994}
\maketitle
\begin{abstract}
Chubukov's proposal concerning the possibility of a nondimerized quantum
nematic phase in the ground-state phase diagram of the
bilinear-biquadratic spin-1 chain is studied numerically. Our results do
not support the existence of this phase, but they rather indicate a direct
transition from the ferromagnetic into the dimerized phase.
\end{abstract}
\bigskip
\section{Introduction}

The general isotropic spin-1 model with nearest neighbor interactions
on a $d$-dimensional lattice is described by the Hamiltonian
\begin{eqnarray}
H &=& \sum_{<i,j>} h_{i,j}
  = \sum_{<i,j>} \left[ \cos\theta\ (\mbox{\boldmath $S$}_{i}\cdot
  \mbox{\boldmath $S$}_{j}) +\sin\theta\ (\mbox{\boldmath $S$}_{i}\cdot
  \mbox{\boldmath $S$}_{j})^2
  \right]\;,                              \label{hami}
\end{eqnarray}
where the parameter $\theta$ governs the ratio of the bilinear and
biquadratic terms. This model has been the subject of several studies in
the last decade
\cite{Papa,Solyom,Parkinson,Klumper,Chub90,Chub91,FaSo1,FaSo2,Xiang,Xian1,Xian}.
The existence of a variety of phases has been demonstrated.
The complete characterization of these phases and of the phase transitions
between them is, however, not yet entirely solved.

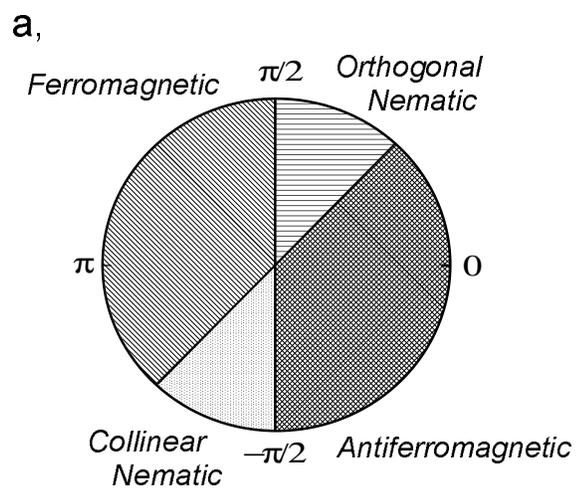
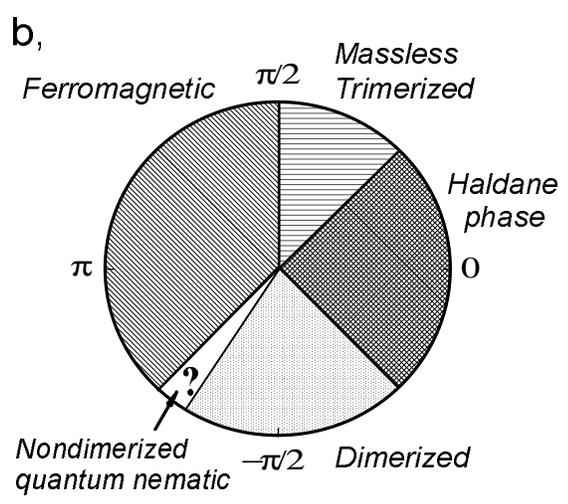
\begin{figure}
\caption{Ground-state phase diagram of the bilinear-biquadratic spin-1 model.
(a) Semiclassical phases without quantum fluctuations ($d\ge 2$).
(b) Quantum phases ($d=1$).}
\label{fig:phasedia}
\end{figure}

For $d\ge 2$ the semiclassical approximation gives a rather good
description of the ground state \cite{Papa}. The zero temperature phase
diagram obtained in this way is sketched in Fig.\ \ref{fig:phasedia}(a). It
consists of four different phases. The usual ferromagnetic ($\pi/2<\theta
<5\pi/4$) and antiferromagnetic ($-\pi/2<\theta <\pi/4$) phases are
separated on both sides by rather exotic phases, where {\em collinear} or
{\em orthogonal nematic} ordering appears. In all phases the original
$SU(2)$ symmetry of the model is spontaneously broken. The order can be
characterized by different long range order parameters. In the nematic
phases, e.g., the expectation value of the spin operators vanishes,
\begin{equation}
   \langle S_i^{\alpha} \rangle =0\;,\qquad \alpha=x,y,z\;,
\end{equation}
while some of the quadrupole operators have finite expectation values with
\begin{equation}
    \langle (S_i^x)^2 \rangle =  \langle (S_i^y)^2 \rangle  \ne
   \langle (S_i^z)^2 \rangle\;.
\end{equation}

Quantum effects, that appear most strongly in one dimension (1D), can
drastically modify this phase diagram. Coleman's theorem \cite{Coleman}
(the quantum analog of the Mermin-Wagner theorem) states that, unless the
order parameter is a conserved quantity, quantum fluctuations will restore
the continuous symmetry of the Hamiltonian. This means that apart from the
ferromagnetic state, whose domain of stability is the same in the
classical and quantum calculations, in 1D the ground state of the model in
Eq.~(\ref{hami}) should be disordered for any other value of  $\theta$. Within
this quantum disordered regime, however, there could  still be several
phases. The semiclassical phase diagram in Fig.\ \ref{fig:phasedia}(a) should
therefore be replaced by a rather different {\em quantum} phase  diagram.

The quantum analog of the classical N\'eel state is the Haldane phase
\cite{Hal}, where dynamic mass generation leads to a unique disordered
ground state with hidden order \cite{denNijs}. There is a finite gap
(Haldane gap) to the excitations. This phase is thought to exist in the region
$-\pi/4 \leq \theta \leq \pi/4$.

The orthogonal nematic phase in $\pi/4 \leq \theta \leq \pi/2$ is replaced
by a tripled periodic phase. There are strong numerical indications
\cite{FaSo1} that this phase is critical due to three soft modes at
$k=0$ and $\pm 2\pi/3$, as is the case in the integrable Lai-Sutherland
\cite{Lai} model corresponding to $\theta=\pi/4$.

Far less understood is the behavior on the other side of the
ferromagnetic regime. In a recent series of papers Chubukov
\cite{Chub90,Chub91} has shown by using perturbational calculation on a
specially chosen ``vacuum'' state with nematic long range order and by
applying the standard renormalization group method, that in 1D in the vicinity
of the ferromagnetic instability point quantum fluctuations restore the
$SU(2)$ symmetry, the ground state remains unique, and the interaction between
the Goldstone bosons leads to dynamic mass generation. The opening of the gap
is exponentially slow as $\theta$ departs from $\theta_{\rm F}$.

Chubukov \cite{Chub91} has emphasized that this disordered phase is
different from the Haldane phase. In fact, the two phases are separated by
a dimerized phase in which translational invariance is broken.
This dimerized phase is believed to exist for $\theta < -\pi/4$. The
spontaneous breakdown of the translational invariance leads to a pair of
singlet ground states and a finite excitation
gap above it. This behavior has been proven rigorously at $\theta=-\pi/2$,
where the system is again integrable and hence the spectrum can be
determined exactly \cite{Parkinson,Klumper}.
The value of the gap at this point is
0.173178, a rather small value, while the correlation length is large,
about 21 lattice units \cite{Klumper}. This explains the early numerical
difficulties in identifying this phase \cite{Solyom}.


As a further support for the existence of the new disordered phase Chubukov
\cite{Chub91} has pointed out that at $\theta_{\rm F}\equiv -3\pi/4$,
where the ferromagnetic state becomes unstable, the magnon spectrum has no
soft mode at $k = \pi$. Neither is there any interesting feature at
$k = \pi$ in the spectrum of bound states of two, three, etc.,
magnons. There does not seem to exist any mode that could drive a direct
transition from the ferromagnet into a dimerized state. In fact, his study
of the effect of quantum fluctuations on the dimerized state, using
appropriate bosonization, led him to conclude that the stability
region of the dimerized phase is $\theta_c \leq \theta \leq -\pi/4$, where
$\theta_c \approx -.7\pi$, thus the disordered phase could exist in
the range $-3\pi/4 \leq \theta \leq \theta_c $.

In what follows this hypothesized phase will be referred to as the {\em
nondimerized quantum nematic} or {\em Chubukov's phase}. The phase diagram
expected to be valid in the 1D case is summarized in
Fig.\ \ref{fig:phasedia}(b).

The main goal of the present paper is to find numerical evidence for the
possible existence of Chubukov's phase. For this purpose we studied
finite spin chains for various values of $\theta$. Beside the
usual L\'anczos diagonalization method, which allowed us to study chains up
to $L=16$ sites with periodic boundary condition, we used White's density
matrix renormalization group method (DMRG) \cite{White} with open
boundary condition up to $L=48$. The data were analyzed by finite-size
scaling techniques.

Since Chubukov's proposed phase is supposed to have a nondegenerate singlet
ground state and unbroken translational symmetry, the transition to this
phase from the dimerized phase can be located either by studying the ground
state degeneracy, the appearance of soft modes at the transition point, or
the behaviour of the dimerization order parameter. Our results for these
quantities will be presented subsequently in the next three sections. Section
V. contains some concluding remarks.

\section{Study of the ground-state degeneracy}

One possibility to see the difference between the dimerized and the
nondimerized quantum nematic phases would be to study the degeneracy of
the infinite volume ground state.

For a finite system with even number of sites $L$ the ground state is unique
for any $\theta > \theta_{\rm F}$; it is an $S_T=0$ spin singlet state.
If periodic boundary condition is used, this state has momentum $k=0$. In
the dimerized phase there has to be another singlet state with momentum
$k=\pi$ with an energy, which, for long enough chains, is exponentially close
to the ground-state energy. This should not be the case, however, for the
nondimerized quantum nematic phase, where the ground state is expected to be
nondegenerate, even in the thermodynamic limit. Thus a gap has to
be opened between the lowest singlet $k=0$ and $k=\pi$ states. Therefore,
we first studied the finite-size scaling behaviour of this gap.

The gaps were calculated by the L\'anczos diagonalization method for chains
with $L= 8, 10, \dots 16$ sites. It is worth pointing out why we could not
use the DMRG method to study this degeneracy problem, although it would
have allowed us to consider much longer chains. Exactly at the point
$\theta_{\rm F}$ the model has an extra $SU(3)$ symmetry and the
Hamiltonian can be written as a (negative) sum of permutation operators.
Any totally symmetric state is a ground state. The ground-state sector is
highly degenerate and belongs to the $D(L,0)$ symmetric representation of
$SU(3)$. This implies that for any finite even chain length $L$ the
ground-state sector consists of the  $S_T=L,L-2,\dots,2,0$ spin
multiplets, each once. Away from $\theta_{\rm F}$  this degeneracy is
lifted, the spin singlet becomes the ground state for  $\theta>-3\pi/4$.
However, in the interesting range of $\theta$ the other states remain
still rather close in energy. For chains with $L=48$ sites many of them
lie still lower than the relevant lowest $k=\pi$ level. They could be
separated by looking at their momentum, since we are interested in a
$k=\pi$ level, while the others have momentum $k=0$. Unfortunately the
DMRG method does not allow to work in a definite momentum sector, and too
many levels would have to be considered. Therefore, using this method, the
interesting gap cannot be computed with a reasonable precision.

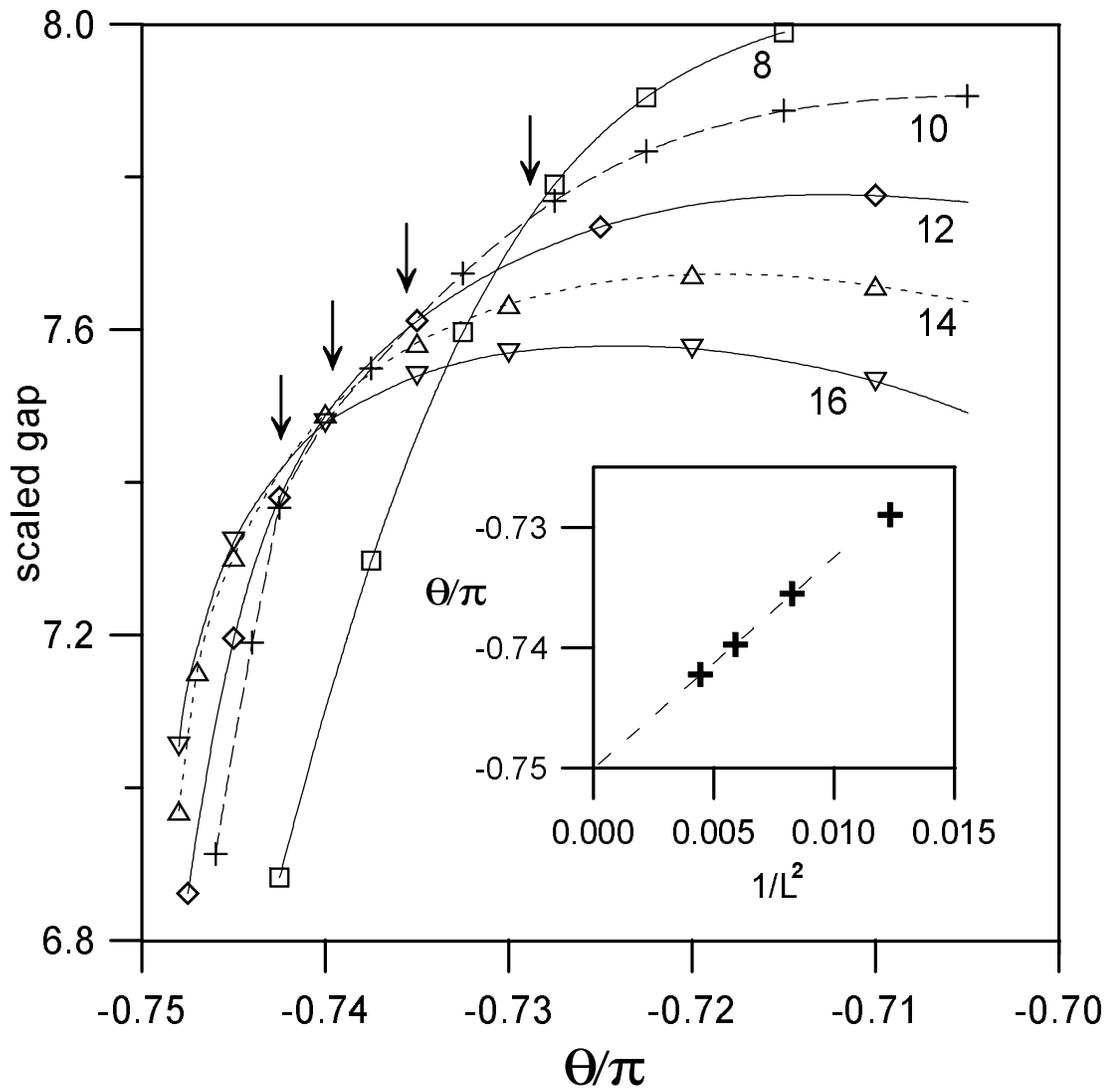
\begin{figure}
\caption{Scaled gap, $L(E_{\pi}-E_{\rm GS})$, between the lowest $k=\pi$
level, and the ground state having momentum $k=0$, as a function of
$\theta$ for chains with $L=8,10,\dots,16$ sites.
Arrows indicate to the crossing point of curves belonging to chain
lengths $L$ and $L+2$. Inset shows the $L^{-2}$ scaling behavior of the
crossing points.}
\label{fig:pbc}
\end{figure}

The scaled gaps for various chain lengths, ie.\ the gaps multiplied by the
number of bonds in the chain are plotted in Fig.\ \ref{fig:pbc}. Close to the
point $\theta_{\rm F}=-3\pi/4$ the scaled gap increases monotonically with
increasing chain length, while further away it decreases. In a critical
model the scaled gap should asymptotically be independent of the chain
length. Therefore, according to the standard procedure the transition
point $\theta_c$ could be located by looking at the crossing points
between curves belonging to chain lengths $L$ and $L+2$. They are marked
by arrows and also shown in the inset of Fig.\ \ref{fig:pbc}. As the chain
length increases, the crossing points scale towards the ferromagnetic
transition point. In fact, as the inset shows, the points fit well to a
straight line on the $L^{-2}$ scale, giving $\theta_c=-3\pi/4=\theta_{\rm F}$.
This would mean that the gap vanishes everywhere above $\theta_{\rm F}$
and Chubukov's phase does not exist.

Due to the limitations of the numerical calculations we cannot exclude
the possibility that the crossing points converge to a
$\theta_c > \theta_{\rm F}$. However, the $\theta$ range where the quantum
nematic phase exists, must then be very narrow with $\theta_c\alt -.74\pi$.

\section{Search for soft modes}

According to Chubukov's proposal the transition at $\theta_c$ between the
nondimerized quantum nematic and the dimerized phases belongs to the Ising
universality class. The lifting of the twofold degeneracy should be
accompanied by the appearance of a soft mode and the vanishing of the gap.
Therefore we have studied the behavior of the gap between the ground state
and the lowest excited state. In the whole interesting region of $\theta$
this lowest excited state was found to have total spin $S_T=2$.

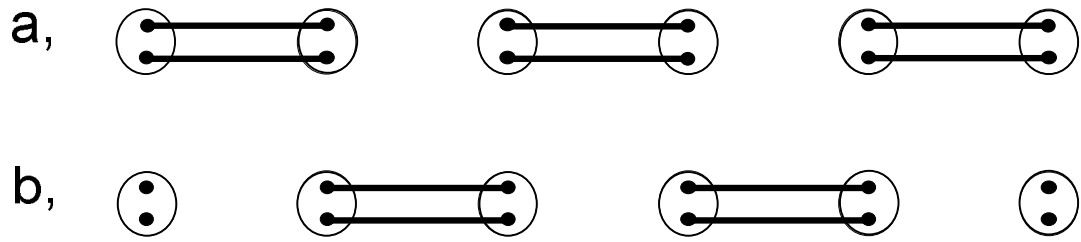
\begin{figure}
\caption{Typical dimerized valence bond configurations with (a)
strong and (b) weak bonds at the ends.}
\label{fig:dimerconf}
\end{figure}

For the calculation of this gap the abovementioned objection does not
hold, and we could apply the DMRG method. Following the usual procedure,
the open boundary condition was used, since it gives much better results than
the periodic one. Now the ground state is expected to be unique even in the
dimerized phase, if the number of sites $L$ is even. This is easily understood
by comparing the two dimerized valence-bond configurations, depicted in
Fig.\ \ref{fig:dimerconf}, which are believed to be good variational states
in the dimerized region. Since the bond-strength oscillates in a dimerized
state, clearly the configuration which has low-energy bonds at the ends has
lower energy than the other one with high-energy bonds at the ends. (For odd
$L$ the two simple dimerized configurations would have the same energy,
therefore they resonate and the
real ground state is a state which has low-energy bonds at both ends and a
moving "domain wall" somewhere in-between \cite{White}.
In this case the ground state
cannot be considered as a real vacuum, since it contains already one
quasi-particle. This fact would make it extremely difficult to draw
conclusions from results on chains with odd $L$, therefore we have not studied
such chains.)

The longest chain we could reasonably study had $L=48$ sites. The
computed value of the gap is believed to be precise to at least five
digits, when the maximum number of block states kept during the
renormalization was $m=160$.

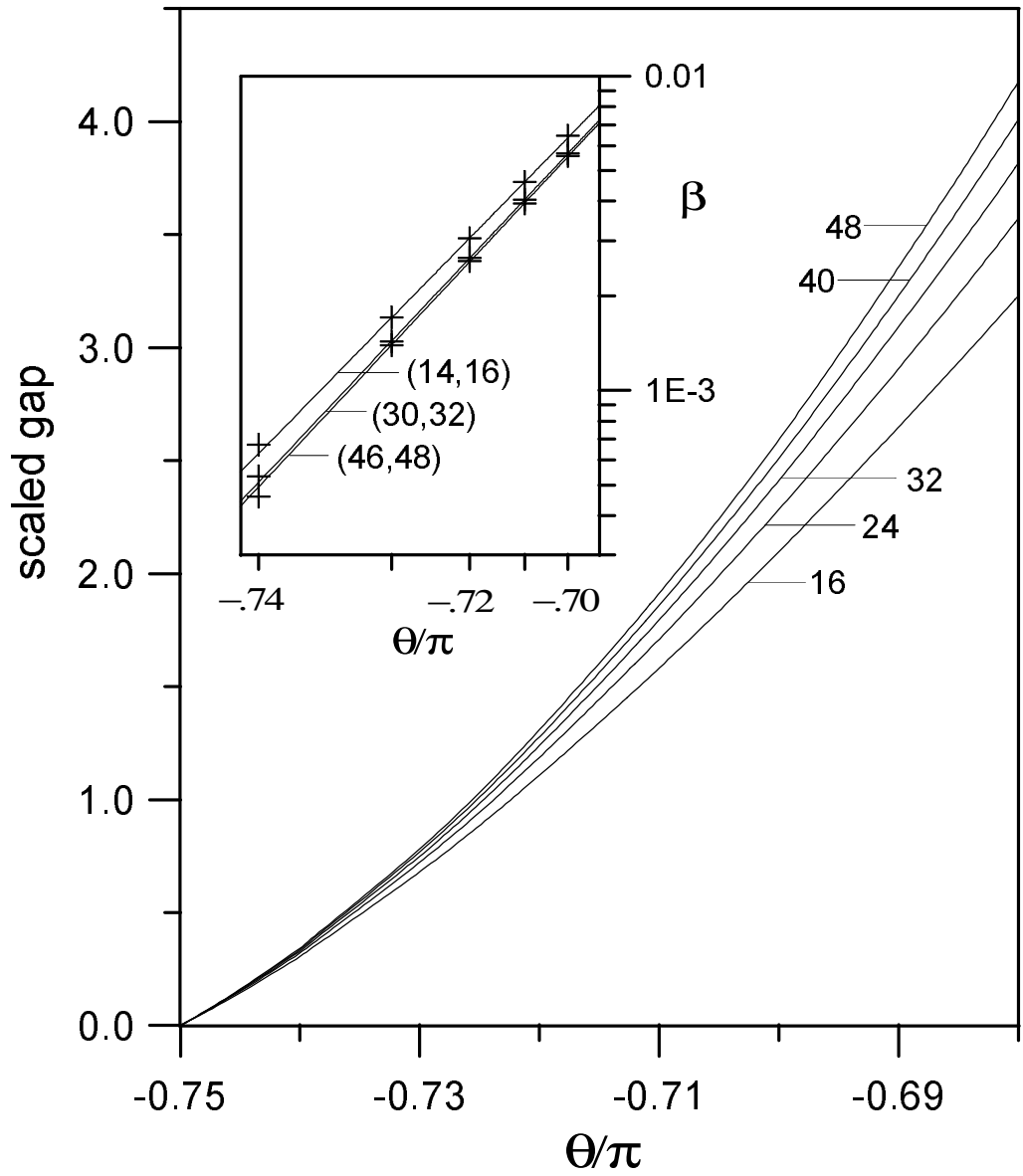
\begin{figure}
\caption{Scaled gap $(L-1)(E_1-E_{\rm GS})$ vs $\theta$, to the lowest
excited state for chains with $L=16,24,\cdots,48$ sites.
Inset shows the Callan-Symanzik $\beta$-function computed from
pairs of chains with length $(L_1,L_2)$ on a log-log scale.}
\label{fig:beta}
\end{figure}

Figure \ref{fig:beta} shows the scaled value of the gap from the ground state
to the lowest $S_T=2$ state in the $-.75\pi<\theta<-.68\pi$ interval. This
gap vanishes exactly even in finite systems at $\theta_{\rm F}=-3\pi/4$
due to the extra $SU(3)$ symmetry, and is extremely small near this
point. Above $\theta_{\rm F}$, however, the scaled gap increases,
indicating the opening of the mass gap in the whole studied interval. In
order to analyze quantitatively the opening of this singlet-quintuplet
gap, we studied the finite lattice approximants of the Callan-Symanzik
$\beta$-function \cite{Roomany,FaSo2}
\begin{equation}
    \beta_{L_1,L_2}(\theta)=
    {\ln [L_2\Delta(\theta,L_2)/L_1\Delta(\theta,L_1)]\over
    \ln (L_2/L_1) {1\over 2} {\partial\over\partial\theta} \ln
             [\Delta(\theta,L_1)\Delta(\theta,L_2)] }\;.\label{beta2}
\end{equation}

$\beta_{L_1,L_2}(\theta)$ depends rather weakly on $L_1$ and $L_2$
(see the inset of Fig.\ \ref{fig:beta}). This allowed us to determine the
opening of the gap at the edge of the ferromagnetic regime rather
convincingly. The $\beta$-function increases above
$\theta_{\rm F}=-3\pi/4$ with a power law
\begin{equation}
    \beta(\theta)=
    -{1\over c\sigma} (\theta-\theta_F)^{1+\sigma} \;, \qquad
                \theta > \theta_{\rm F}\;,
      \label{betafit}
\end{equation}
giving a straight line on the log-log plot, like in a
Kosterlitz-Thouless transition. The best fit was obtained
with $c=3.9\pm0.4$, $\sigma=.51\pm 0.03$. This means
that the gap opens exponentially as
\begin{equation}
  \Delta(\theta)=
  {\rm const}\cdot \exp[-c (\theta-\theta_F)^{-\sigma}]\;, \qquad
                \theta >\theta_{\rm F}\;.
\end{equation}
This is in agreement with Chubukov's result, which suggested an
exponentially slow opening of the gap \cite{Chub90}. Note that although
this form resembles that of the Kosterlitz-Thouless transition, we think that
the phase transition here is not really a KT transition but a first order
one because the transition from the ferromagnetic to this state is due to
the crossing of the $S_T=L$ and $S_T =0$ levels, which are the ground
states on the two sides, respectively.

For $\theta>\theta_{\rm F}$, for the chain lengths we could use, there
did not appear any sign of further level crossings. We did not find any
level that would soften at a $\theta_c \ne \theta_{\rm F}$, ie.\ we did not
find any  trace of an additional phase transition. It should be noted,
however,  that the DMRG method gives precise energies only for those
states that are targeted from the very beginning of the iteration. The
procedure can easily miss a level, thus leading to false conclusion, if its
energy comes down into the interesting energy range for long system sizes
only, and hence it is not targeted at the beginning of the algorithm.

\section{Study of the dimerization order parameter}

An alternative way to study phase transitions is by analyzing order
parameters and correlation functions. They require the knowledge of the
ground-state wave function only.

Even though both the dimerized phase and the nondimerized quantum nematic
phase are disordered, an order parameter can be defined in the dimerized
phase by using the fact that the translational invariance is spontaneously
broken. The dimerization order parameter is usually defined as
\begin{equation}
D \equiv |\langle \mbox{\boldmath $S$}_{i-1}\cdot \mbox{\boldmath $S$}_{i}
  - \mbox{\boldmath $S$}_{i}\cdot \mbox{\boldmath $S$}_{i+1}\rangle | \;.
\end{equation}
In Ref.\ \cite{Xian} Xian pointed out that for the pure biquadratic model
a more proper definition would be the bond-strength
oscillation in the ground state.
Following this suggestion we define the dimerization order parameter as
\begin{equation}
D= |\langle h_{i-1,i}-h_{i,i+1} \rangle| \:,
\end{equation}
where $h_{i,j}$ is the local Hamiltonian of Eq.\ (\ref{hami}), and the
expectation value is taken in the ground state of an infinite
chain whose ends are subject to open boundary condition.

In case of finite chains, let us denote the energy difference of
neighboring bonds in the middle of the open chain with length $L$ by
\begin{equation}
D(L)= |\langle h_{L/2,L/2+1}-h_{L/2+1,L/2+2} \rangle |\:.
\end{equation}
The (infinite volume) dimerization order parameter is then
\begin{equation}
D= \lim_{L\to\infty} D(L).
\end{equation}

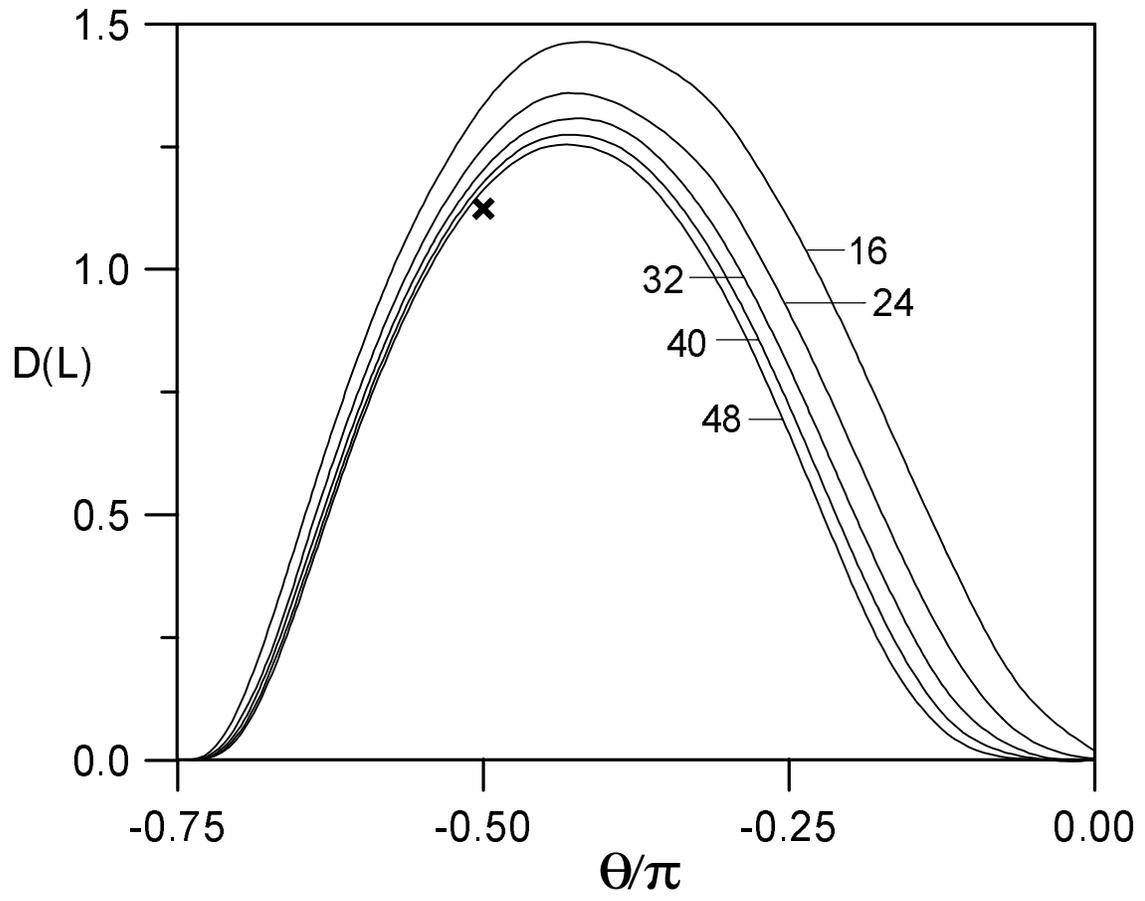
\begin{figure}
\caption{Dimerization order parameter $D(L)$ vs $\theta$ for chains
with $L=16,24,\cdots,48$ sites. The cross shows Xian's exact
value for $D$ at $\theta=-\pi/2$.}
\label{fig:dimer}
\end{figure}

$D(L)$ is shown in Fig.\ \ref{fig:dimer} in the range $\theta_{\rm
F}\le\theta\le 0$. It is finite also for $\theta \geq -\pi/4$, where no
dimerization is expected. It vanishes at the VBS point $\theta=\arctan 1/3$
\cite{Affleck} only, where the ground state has a simple form and
$D(L)\equiv 0$ for all $L$.

The finiteness of $D(L)$ can be understood by observing
that the bond strength is generally not uniform in an open chain near
the ends but alternate between strong and weak bonds, even if in the
thermodynamic limit the system is not dimerized \cite{White}.
As we move towards the
middle of the chain this oscillation gradually decreases. In a noncritical
model this decay is exponential. In leading order $D(L)$ is expected to
vary as
\begin{equation}
D(L) = D + c \exp (-L/\xi)\;,  \label{offcrit}
\end{equation}
where $c$ is a constant and $\xi$ is a kind of a correlation length.
When the model has a nondimerized thermodynamic limit, $D=0$.

In case of critical behavior, however, $D(L)$ scales to zero
as a power law, ie.\
\begin{equation}
D(L) =  c L^{-\alpha}\;, \label{crit}
\end{equation}
where $\alpha$ is a relevant surface exponent. Using this property, a
systematic scaling analysis of our data for $D(L)$ allows to determine $D$
for various values of $\theta$.

As a first test we computed $D(L)$ at $\theta=-.5\pi$, where the exact result
for $D$ is available. Fitting the values of $D$, $c$ and $\xi$ of
Eq.\ (\ref{offcrit}) to the last three points, $L=40,44$ and $48$, we got
$D=1.1288$ which is quite close to Xian's \cite{Xian} exact value
$D_{\rm exact}=1.1243$, when it is properly rescaled according to
our definition of $D$. Our value for the correlation length is
$\xi=19.93$, which compares well with the exact value for bulk
correlation length $\xi_{\rm exact}=21.0728$ \cite{Klumper}.

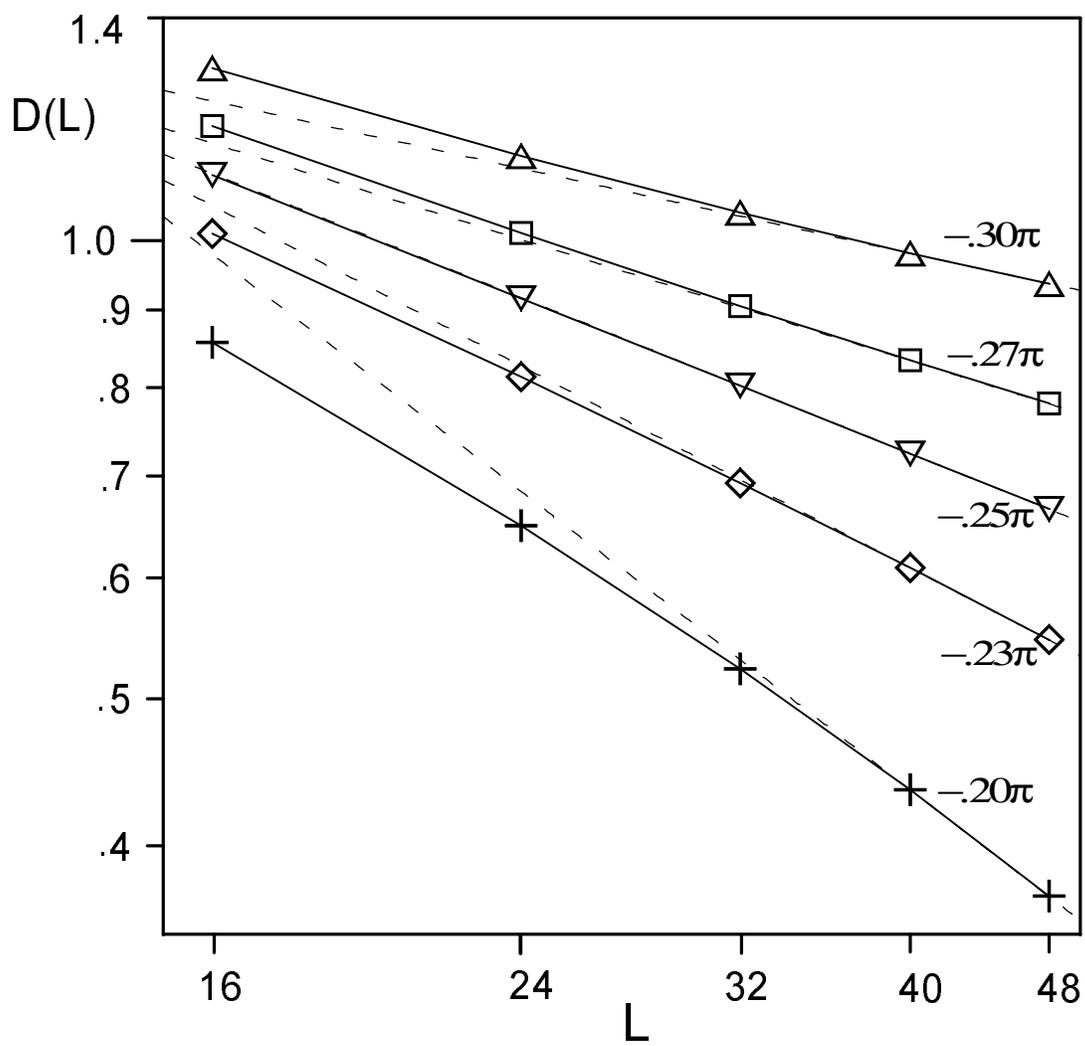
\begin{figure}
\caption{Log-log plot of the dimerization order parameter $D(L)$ vs the
chain length $L$ for different values of $\theta$ near
$\theta=-\pi/4$.
Dotted lines are guides to the eye. They are straight lines fitted to
the last two points.}
\label{fig:loglogtb}
\end{figure}

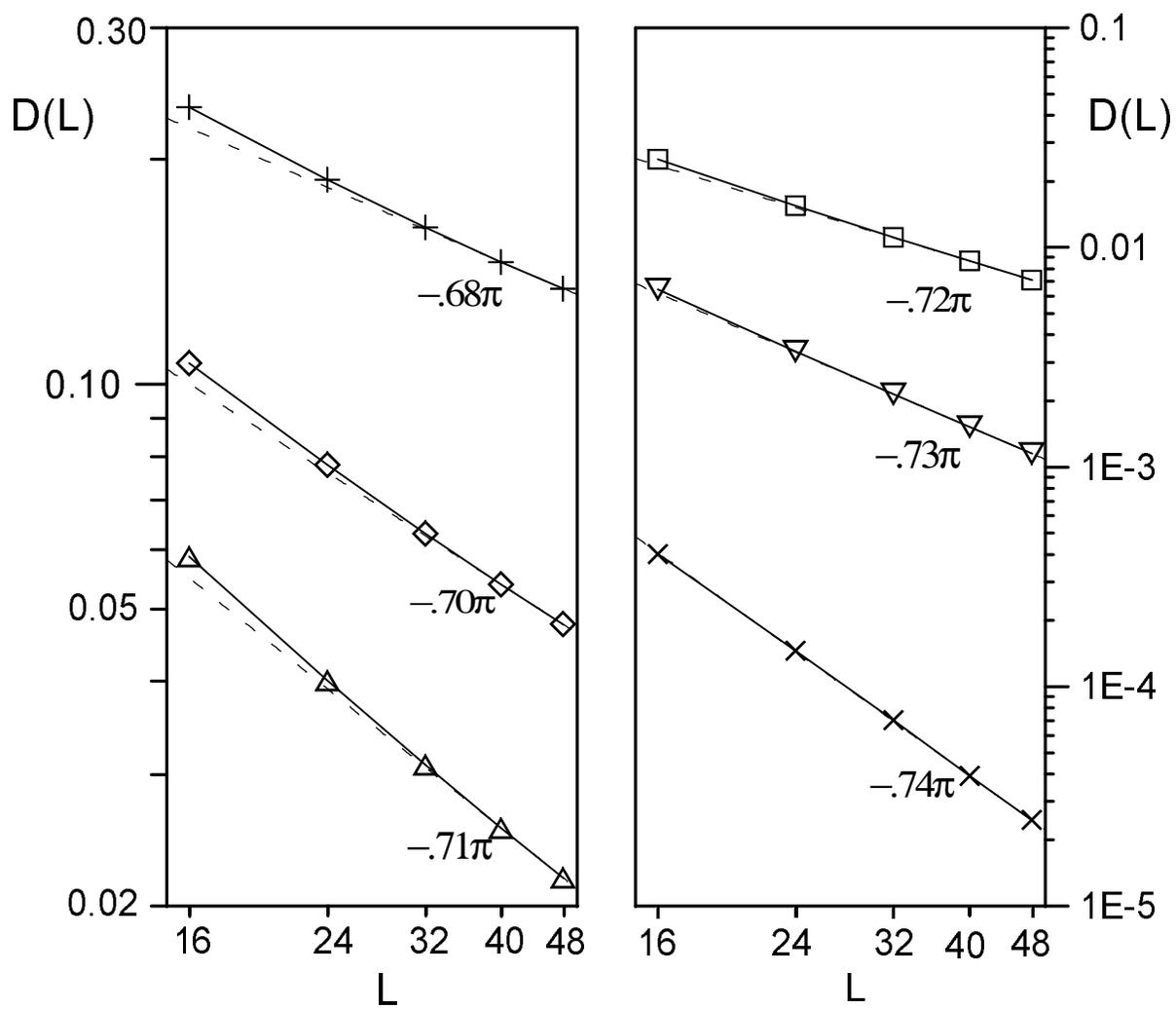
\begin{figure}
\caption{Log-log plot of the dimerization order parameter $D(L)$ vs the
chain length $L$ for different values of $\theta$ near
$\theta_{\rm F}=-3\pi/4$.
Dotted lines are guides to the eye. They are straight lines fitted to
the last two points.}
\label{fig:loglogdi}
\end{figure}

Figures \ref{fig:loglogtb} and \ref{fig:loglogdi} show a log-log plot
of $D(L)$ vs $L$ for various values of $\theta$ near $\theta=-\pi/4$
and $\theta_{\rm F}$, respectively. The transition between the
dimerized phase and the Haldane phase is easily spotted: for
$\theta <-\pi/4$ the curves have an upward curvature indicating a
finite value for $D$, while for $\theta >-\pi/4$ the curvature is
in the opposite direction implying $D=0$. At the critical point,
where the asymptotic scaling is expected to be of the form of Eq.\
(\ref{crit}), the points should lie on a straight line. From this analysis
we obtain $\theta=(-.25 \pm .01)\pi$ for the transition point, as expected.

In contrast to this behavior, all the curves have a slight upward curvature
for $\theta_{\rm F} < \theta<  -.68\pi$, although this curvature is almost
zero in the close vicinity of $\theta_{\rm F}$. This implies a very
small but finite dimerization in this region as well.  Our results are,
however, more uncertain
in this region than near $\theta = -\pi /4$. Since $D(L)$ is small,
the relative accuracy of the method is not very satisfactory, even when
$m=160$ basis states are kept during the renormalization. There
is also a possibility that we are not yet in the asymptotic regime.
Nevertheless, these results seem to indicate that most likely $D$ remains
finite down to $\theta_{\rm F}$. Like the gap, its value becomes
exponentially small near the transition point.

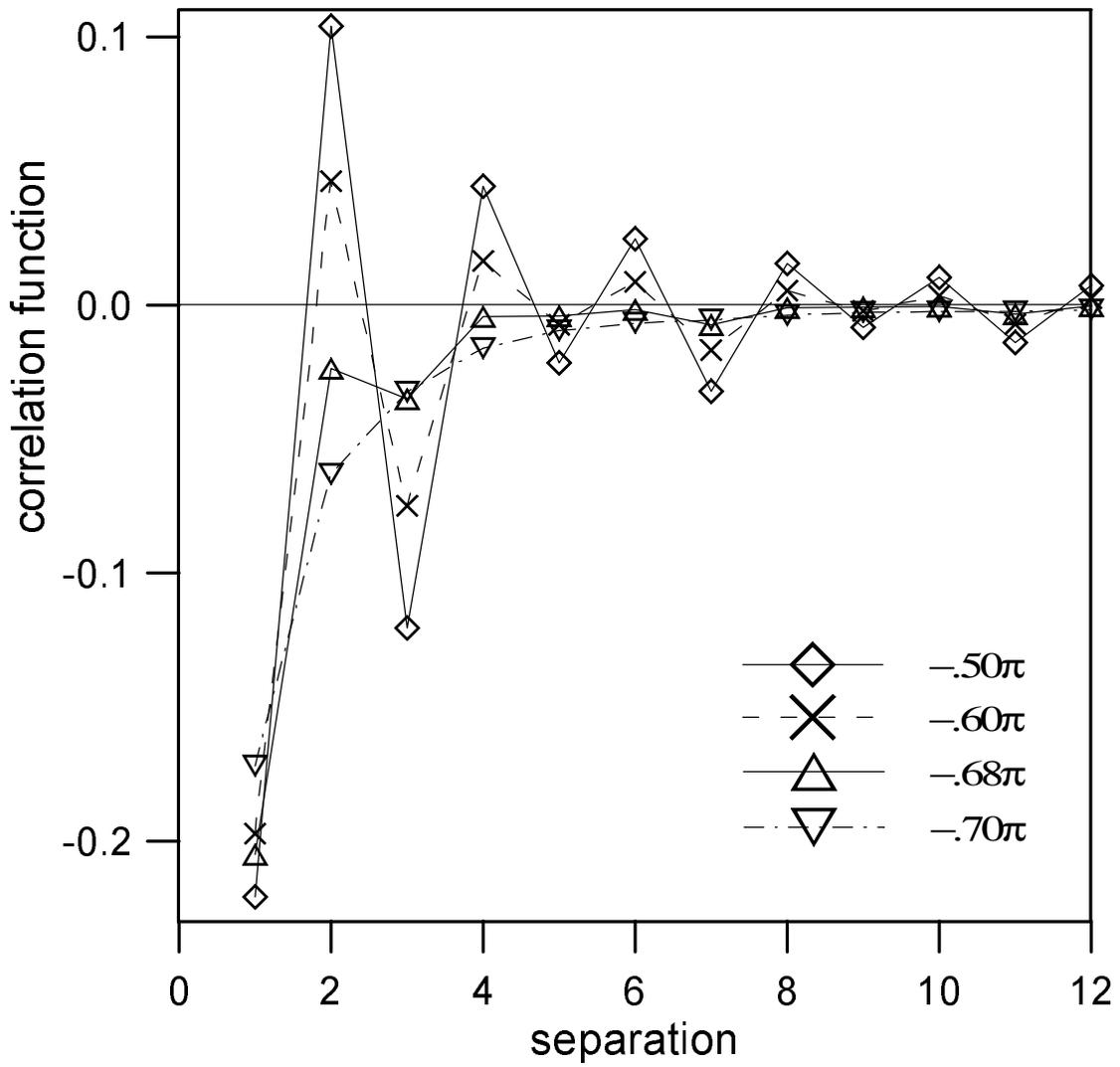
\begin{figure}
\caption{Ground-state correlation function $\langle S^z_0 S^z_l
\rangle$ measured in a chain with $L=48$ sites
as a function of the separation $l$ for different values of $\theta$.}
\label{fig:corr}
\end{figure}

Finally in Fig.\ \ref{fig:corr} we present the two-point correlation function
$\langle S_0^zS_l^z \rangle $  for three different values of $\theta$. Well
inside the dimerized region the correlation function alternates in sign
strongly. It is positive for even $l$, and negative for odd $l$.
For smaller values of $\theta$, however, this behavior changes
drastically. At $\theta=-.68\pi$ already all correlations are found to be
negative and this property remains valid down to $\theta_{\rm F}$. There
appears to be no sign of period doubling, just as it was proposed for the
nondimerized quantum nematic phase. Although we do not fully understand
the reason of this change, we think that the presence of
dimerization should not necessarily mean an alternation in the sign
of the correlation function, and thus the observed
behavior cannot be viewed as a direct evidence for the appearance
of a nondimerized regime.

\section{Conclusions}

In the present paper we studied the ground-state phase diagram of the
bilinear-biquadratic spin-1 chain near the ferromagnetic instability,
where Chubukov proposed the existence of a nondimerized quantum nematic phase.
We considered four independent quantities. First we looked at the
degeneracy of the ground state, ie.\ whether the ground state is unique or
doubly degenerate. Then we determined the opening of the gap. Finally we
studied the behavior of the dimerization order parameter and the ground-state
correlation function. We did not find any real evidence
in any of these quantities for the existence of the hypothesized
nondimerized quantum nematic phase.

Our results are in better agreement with the assumption that the
dimerization appears exactly at the point where ferromagnetism becomes
unstable, ie.\ at $\theta_{\rm F}=-3\pi/4$. The gap opening is well
described by a Kosterlitz-Thouless-like form, although the transition is
expected to be of first order.

The main difficulty we had to face in our numerical work was the extreme
smallness of the gaps and order parameters in the region where the
nondimerized quantum nematic phase was proposed to exist. Due to the
restrictions in the numerical calculations and the uncertainties in the
extrapolation procedure, our conclusion cannot be definitive. We cannot
exclude the possibility that the phase proposed by Chubukov does appear in
a very narrow region, say $-.75\pi \leq \theta \alt -.74\pi$.
To resolve this question satisfactorily calculations on even
longer chains would be needed.

\section{Acknowledgments}

This research was supported in part by the Hungarian Research Fund
(OTKA) Grant Nos.\ T4473 and 2979.  The authors acknowledge
gratefully the enlightening discussions with A.\ V.\ Chubukov and
K.-H.\ M\"utter.

\newpage

\end{document}